\documentclass{article}

\usepackage{arxiv}

\usepackage[utf8]{inputenc} 
\usepackage[T1]{fontenc}    
\usepackage{hyperref}       
\usepackage{url}            
\usepackage{booktabs}       
\usepackage{amsfonts}       
\usepackage{amsmath}
\usepackage{nicefrac}       
\usepackage{microtype}      
\usepackage{graphicx}
\usepackage{natbib}
\usepackage{doi}

\title{Integration of solid-state nanopore arrays via dry bonding to photostructured microfluidic networks}

\author{ \href{https://orcid.org/0000-0003-3200-3217}{\includegraphics[scale=0.06]{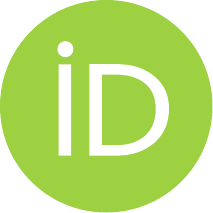}\hspace{1mm}Peter D. Jones} \\
	NMI Natural and Medical Sciences Institute \\at the University of Tübingen\\
	72770 Reutlingen\\
	Germany\\
	\texttt{peter.jones@nmi.de} \\
	\And
	\href{https://orcid.org/0000-0001-6958-8218}{\includegraphics[scale=0.06]{orcid.pdf}\hspace{1mm}Michael Mierzejewski} \\
	NMI Natural and Medical Sciences Institute \\at the University of Tübingen\\
	72770 Reutlingen\\
	Germany\\
	\texttt{michael.mierzejewski@nmi.de} 
}

\renewcommand{\shorttitle}{Solid-state nanopore arrays via dry bonding to microfluidics}

\hypersetup{
pdftitle={Integration of solid-state nanopore arrays via dry bonding to photostructured microfluidic networks},
pdfsubject={cond-mat.mes-hall, physics.app-ph},
pdfauthor={Peter D. Jones, Michael Mierzejewski},
pdfkeywords={},
}

\begin{document}
\maketitle

\begin{abstract}

The integration and parallelization of nanopore sensors are essential for improving the throughput of nanopore measurements. Solid-state nanopores traditionally have been used in isolation, which prevents the realization of their full potential in applications. In this study, we present the microfluidic integration of an array of 30 nanopores, which, to our knowledge, is the highest number reported to date. Our microfluidic network was fabricated using high-resolution epoxy photoresists, and the solid-state membranes were bonded through a dry process using complementary surface chemistries. We successfully measured integrated nanopores using external electrodes. This paper discusses the limitations of our methods, particularly concerning microfluidic interfacing and scaling to higher channel counts. Additionally, we present theoretical analysis of current blockades and noise in integrated nanopores, predicting that maintaining low series resistance between the nanopore and electrode is crucial for resolving short events.

\end{abstract}


\section{Introduction}\label{introduction}

Nanopore sensors have an enormous potential. Commercial success has been
achieved with biological nanopores, while promising basic research with
solid-state nanopores remains to be translated into practical use.
Parallelization has proven to be an important factor for efficient
research on biological nanopores and is critical for practical use
\citep{Laszlo2014}. Examples include the hundreds of nanopores per
Oxford Nanopore Technologies flow cell (5~kHz bandwidth) or the sixteen
nanopores in Ionera microelectrode-cavity array (MECA) chips (100~kHz
bandwidth with Nanion systems). In contrast, a key weakness of
solid-state nanopores is their measurement as single pores. Few works
integrate multiple solid-state nanopores (Table~\ref{tab:SotA}).

A primary challenge for arrays of solid-state nanopores is the filling
of the trans compartment with salt solution, while maintaining isolation
between neighboring nanopores \citep{magierowski_nanopore-cmos_2016}. Consider
biological nanopores: the trans compartments are initially wetted during
filling of a common cis compartment. Thereafter, formation of a lipid
bilayer encloses the trans compartments and forms an insulating
membrane. Biological nanopores then spontaneously enter the membrane. In
contrast, solid-state membranes of silicon nitride or other engineering
materials cannot be spontaneously formed in place. The trans
compartments must therefore be filled by another means. Good methods for
wetting the trans compartments of an array of solid-state nanopores
should have the capacity to be scaled to at least hundreds of pores to
compete with biological nanopores, and must be user-friendly to be
viable for envisioned applications.

A pioneering work integrated solid-state nanopores in PDMS microfluidics
by transfer printing of free-standing membranes \citep{Jain2013}.
Other works have similarly bonded membranes supported by a silicon frame
to PDMS microfluidics \citep{Tahvildari2015, Tahvildari2016}. Disadvantages of
PDMS include its poor compatibility with microfabrication processes, its
capacity to absorb small molecules, and its hydrophobicity. PDMS
microfluidics also do not easily allow electrodes near the integrated
nanopores.

These works achieved poor usability, as each nanopore is in its own
microfluidic channel. To simplify this, integration of microfluidic
valves in PDMS has been demonstrated \citep{Tahvildari2016, jain_microfluidic_2017}. Jain et al. discussed how the higher possible density of
nanopores vs. density of electrodes and amplifiers supports the use of
fluidic switching of nanopores. They used a three-bit multiplexer to
address eight channels, demonstrating how microfluidic valves can allow
higher density of nanopores and reduce the number of fluidic
connections. For a given number of nanopores, the number of pneumatic
channels is the binary logarithm. Further development to integrate individual electrodes for each nanopores, use alternative elastomeric materials instead of PDMS, and to increase the number of channels would make this a promising concept.

\begin{table}[]
	\caption{Overview of microfluidic integration of solid-state nanopores.}
    \begin{tabular}{@{}cp{4cm}p{5cm}p{4cm}@{}}
\toprule
Number of \\nanopores & Details                               & Fluidic interfacing                                      & Reference                 \\ \midrule
1                   & Nanopore chip assembled with silicone & Reservoir on both sides                                  & \citet{rosenstein_integrated_2012} \\
1                   & SiN membrane sandwiched between PDMS  & One channel on each side of the membrane                 & \citet{Jain2013}       \\
5                   & Nanopore chip sandwiched between PDMS & Five channels / shared channel                           & \citet{Tahvildari2015} \\
5                   & Nanopore chip sandwiched between PDMS & Five channels with microfluidic valves / shared channel  & \citet{Tahvildari2016} \\
8                   & SiN membrane sandwiched between PDMS  & Eight channels with microfluidic valves / shared channel & \citet{jain_microfluidic_2017}       \\
30                  & Bonded on one side to epoxy channels  & Thirty channels / open reservoir                         & This work                 \\ \bottomrule
    \end{tabular}
	\label{tab:SotA}
\end{table}

Nanopore array fabrication methods must also be developed. While the
smallest nanopores are produced by electron beam milling in a
transmission electron microscope (TEM), this method is poorly suited for
arrays. Ion beam milling in a helium ion microscope (HIM) provides very
small nanopore sizes and better scalability \citep{emmrich_nanopore_2016, xia_rapid_2018}. Controlled breakdown has been demonstrated with
microfluidic integration \citep{Tahvildari2015}. The most scalable
method would likely be reactive-ion etching (RIE) \citep{bai_fabrication_2014, martens_nanopore-fet_2022}. Beyond parallelization, integration can reduce
noise in nanopore sensors \citep{rosenstein_integrated_2012, Jain2013}.

\begin{figure}
    \centering
    \includegraphics{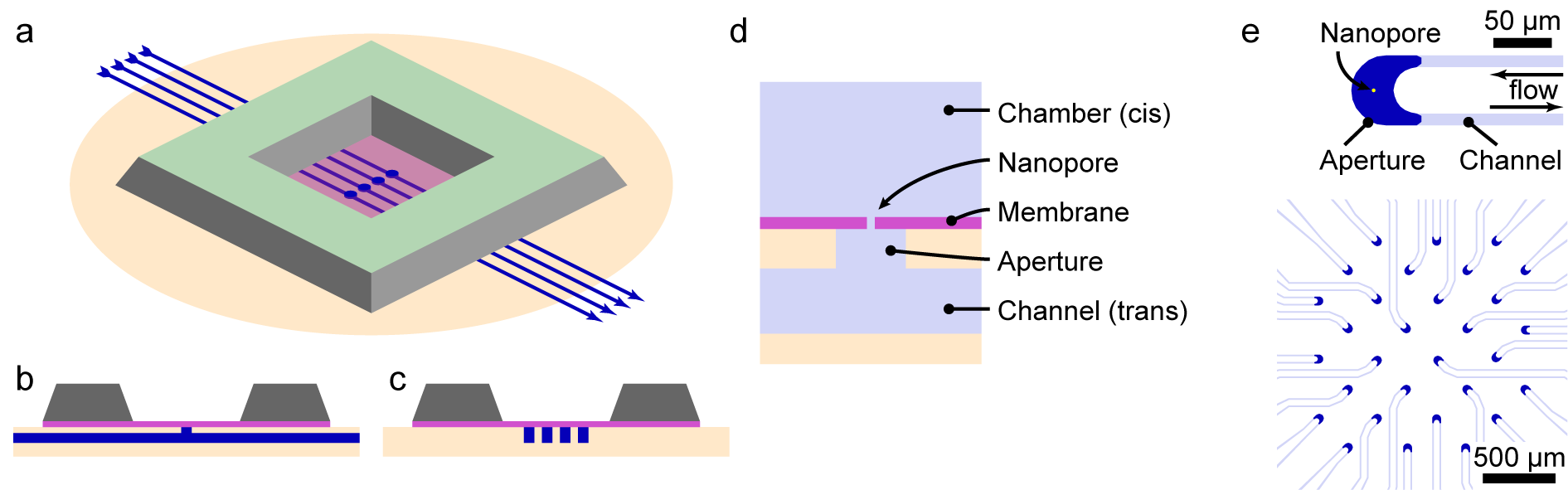}
    \caption{Concept. \textbf{a:}~An isometric illustration showing 
nanopores in individual microfluidic channels. Only four channels are shown for simplicity. A nanopore membrane chip
with a large window is bonded to microfluidic channels. \textbf{b:}~A
cross-section along a channel, showing an aperture in the channel
enclosed by the nanopore membrane. \textbf{c:}~A cross-section
perpendicular to the channels, showing four apertures enclosed by the
nanopore membrane. \textbf{d:}~A magnification of (b), indicating the
cis chamber and trans channel. \textbf{e:} Our design uses an
out-and-back channels (above) to route all inflow and outflow channels
for 30~nanopores in a 2×2~mm² region (below).}
    \label{fig:concept}
\end{figure}

Here, we present a new concept for microfluidic integration of nanopore
arrays in permanent epoxy-based microfluidics (Figure~\ref{fig:concept}). While originally developed towards the application of nanofluidic chemical neurostimulation \citep{jones_nanofluidic_2017}, the concept and fabrication methods are promising for nanopore sensors. Our microfluidic
structures with a thickness below 30~µm are compatible with
substrate-integrated microelectrodes on glass or silicon. We demonstrate
the permanent bonding of solid-state membranes with
30~focused-ion-beam-milled nanopores. Bonding is performed after
independent microfabrication (of electrodes and microfluidics) and
nanofabrication (of nanopores) to maximize process flexibility. We
present electrochemical measurements and calculations demonstrating
potential noise reductions with our integrated arrays.

We furthermore discuss limitations of our nanopore array devices,
including poor user-friendliness (with a custom 60-tube microfluidic
connector). The need for two tubes per nanopore limits scalability. Our
demonstration used large nanopores milled by a gallium beam and measured
with external electrodes, rather than integrated microelectrodes.

Despite these limitations, our methods should enable fabrication of
nanopore arrays if new concepts can solve the microfluidic interfacing
and usability issues. Separation of micro- and nanofabrication enables
compatibility with nanopore fabrication by methods such as milling with
focused electron or ion beams, reactive-ion etching or controlled
breakdown.

\section{Methods}\label{methods}

\subsection{Nanopore milling}\label{nanopore-milling}

Chips (5~mm square) with freestanding SiN\textsubscript{x} membranes 
were purchased from Silson Ltd. (Blisworth, England). Nanopores were
produced by focused-ion-beam (FIB) milling using the gallium beam of an AURIGA CrossBeam
FIB-SEM (Carl Zeiss AG, Oberkochen, Germany). Before milling, membranes
were sputter coated with \textasciitilde10~nm AuPd for conductivity.
Nanopores were milled from the etched side of the nanopore chip to
enable simple alignment to the membrane corners. The gallium beam was
manually focused. Digital beam control held the focused beam (with a
current of 1--20~pA) at a single location for up to 60~s to mill
individual pores. Arrays were milled on 2×2~mm² membranes and stage
movement was used to produce an array over an area of 1.4×1.4~mm².

\subsection{Microfluidic fabrication}\label{microfluidic-fabrication}

\begin{figure}
    \centering
    \includegraphics{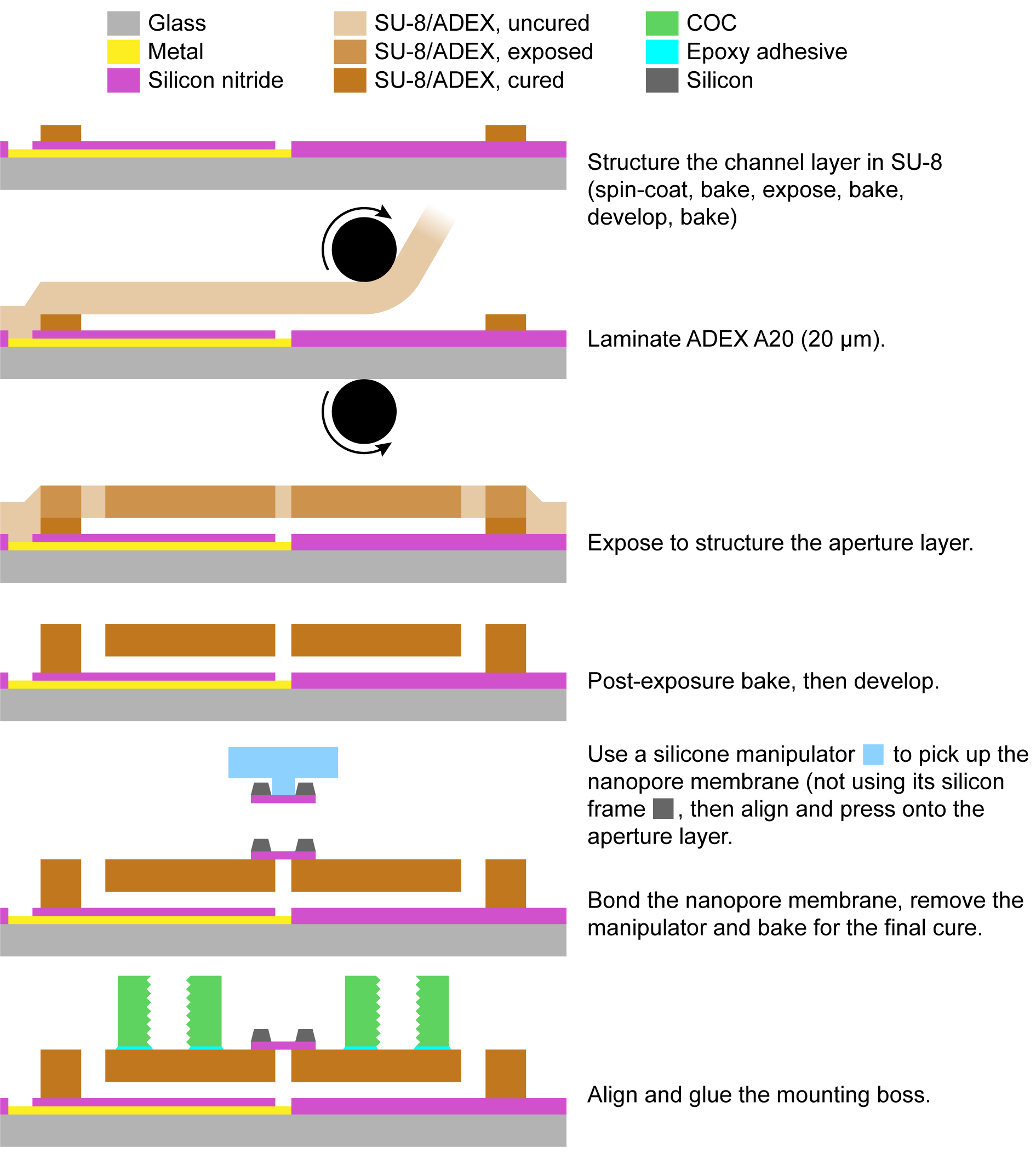}
    \caption{Fabrication scheme. Here, the fabrication of nanopore arrays
is illustrated on a glass-based microelectrode array with a silicon
nitride insulator. The same methods were used on glass substrates with
unstructured silicon nitride coatings and no electrodes.}
    \label{fig:microfab}
\end{figure}

Figure~\ref{fig:microfab} illustrates the microfabrication methods. Microfluidic
structures were produced by photolithography of a first spin-coated 
photoresist layer to define channels and a second dry film photoresist
layer to enclose channels and define apertures. Glass
substrates coated with silicon nitride were prepared by oxygen plasma
treatment (2~min) and baking in an oven at 150~°C for 3~h. The channel
layer of SU-8 was spin-coated to a nominal thickness of 10~µm (10~s at
500~rpm then 30~s at 1000~rpm) then baked on a hotplate for 1~min at
65~°C, 3~min at 95~°C and 1~min at 65~°C. The layer was patterned by UV
exposure (430~mJ/cm², i-line filter, MA6 mask aligner, SÜSS MicroTec
AG). The post-exposure bake was 1~min at 65~°C, 5~min at 95~°C and 1~min
at 65~°C. Substrates were developed in mr-Dev 600 (micro resist
technology GmbH, Berlin, Germany) for 1 min, then rinsed with
isopropanol and dried with nitrogen. Substrates were baked in an oven at
150 °C for 1 h for resistance against later exposure to cyclohexanone.

The aperture layer was formed by lamination of 20~µm-thick ADEX dry film
resist (DJ Microlaminates, Sudbury, MA, USA). The film was laminated at
80~°C and 3~mm/s. The backing foil was removed before patterning by UV
exposure (1000~mJ/cm² with i-line filter). The post-exposure bake was
1~min at 65~°C, 10~min at 85~°C and 1~min at 65~°C. Substrates rested
for 2~h and then were developed in cyclohexanone for 3~min, rinsed with
fresh cyclohexanone, rinsed with isopropanol and blow-dried with
nitrogen with care to clear any solvents from the channels. The
substrates were then ready for bonding with nanopore membranes.

\subsection{Bonding}\label{bonding}

A dry bonding method was developed similar to \citet{Zhang2009}.
We first modified nanopore membrane chips using an amino-based adhesion
promoter as follows. Piranha solution was prepared by adding 1 part 30~\% H2O2 to 3 parts
96\% H2SO4. Chips were submerged for 1~h, during which time the
temperature remained above 100~°C. The surface of silicon nitride
contains silicon oxide \citep{Raider1976} and piranha cleaning forms
silanol groups \citep{acres_molecular_2012}. Piranha also removed the AuPd
layer. Chips were then rinsed with deionized water and blown dry with
nitrogen. Membranes were coated with 3-aminopropyldimethylethoxysilane
(APDMES) by immersion (overnight or longer) in a 1~\% v/v solution in
toluene. Finally, chips were rinsed with toluene, acetone, and
isopropanol, dried with nitrogen, and baked in an oven at 120~°C for
30~min.

Nanopore membranes were bonded on the microfluidic aperture layer by a
flip chip process (Finetech GmbH \& Co. KG, Berlin, Germany). A custom 
silicone manipulator (1.9×1.9~mm²) picked up the membrane directly using
van der Waals forces. We fluorinated the manipulator using 
1H,1H,2H,2H-perfluorooctyltrichlorosilane to prevent chemical bonding to the membrane. The membrane and microfluidic apertures were
optically aligned, then brought into contact. A force of 2--3~N was
applied, and the temperature was increased at 1~°C/s to 100 C, held for
45 min, then decreased at the same rate; the manipulator was lifted from
the membrane after returning to room temperature. Finally, substrates
were hard-baked at 170 to 200 °C for several hours.

\subsection{Microfluidic connectors}\label{microfluidic-connectors}

A connector was produced in a design adapted from \citet{Wagler2015}
with inspiration from linear connectors from Dolomite Microfluidics. The
connector consisted of silicone seals and a clamp, and additionally
required a mounting boss on the microfluidic substrates. The clamp and
mounting boss were produced by CNC-milling in aluminum and cyclic olefin
copolymer (Topas 6015), respectively. The seals were produced by casting
silicone (Sylgard 184) in a custom casting setup produced by CNC
milling. Mounting bosses were glued to the substrates using an epoxy
adhesive.

We used a casting setup like that of Wagler et al. to produce 30-channel
silicone seals. A milled part with holes in a 2~mm pitch held metal needles (0.6~mm, Prym, Stolberg, Germany) to immobilize
silicone tube ports (5~mm long, 0.51~mm inner diameter, VWR
International GmbH, Darmstadt, Germany) in the casting setup. The tubes
were activated by air plasma, the setup was filled with Sylgard 184
mixed at a 10:1 ratio, and the silicone was allowed to cure at room
temperature to avoid shrinkage \citep{Madsen2014}.

\subsection{Hydraulic resistance}\label{hydraulic-resistance}

For cylindrical channels, hydraulic resistance \(R_{\text{H}}\) was
calculated by

\[R_{\text{H}} = \frac{128\eta L}{\pi d^{4}}\]

for a channel length \(L\) and diameter \(d\) and a dynamic viscosity
\(\eta\) (0.89~mPa·s for water). For rectangular channels, the
resistance was estimated as

\[R_{\text{H}} = \frac{12\eta L}{wh^{3}\left( 1 - \frac{0.63h}{w} \right)}\]

which is accurate to within about 13~\% for square channels \citep{Bruus2015}.

To experimentally measure hydraulic resistance, we measured the time required for a meniscus to move a given distance in tubing of known diameter.

\subsection{Measurement of nanopore
arrays}\label{measurement-of-nanopore-arrays}

For each nanopore, 2~µL of phosphate-buffered saline (PBS) was injected
through the inlet tube to fill the microfluidic channel until a droplet
of PBS formed at the tip of the outlet tube. A 1~mL syringe containing
PBS and a trans electrode was brought into contact with the outlet tube
droplet (Figure~\ref{fig:circuit}a). The cis electrode was placed in the common bath
above the nanopore membrane. Electrochemical impedance spectroscopy
(EIS) was performed using a potentiostat (Bio-Logic Science Instruments
SAS, Claix, France). Afterwards, the channel was flushed with 2~µL of
deionized water and finally pressurized air. This process was repeated
for each channel. Flow was controlled by application of 1~bar of air
pressure with a Luer-Lock blunt needle (outer diameter 0.8~mm) connected
to each tube using a short segment of silicone tube. Before measurement,
the nanopore array was treated with air plasma for 2~min to improve
hydrophilicity. All electrodes were Ag/AgCl wire electrodes.

We use the following expression from \citet{Kowalczyk2011a} to relate
conductance $G$ to diameter $d$ and length $l$ of a
nanopore filled with a salt solution with conductivity $\sigma$:

\begin{equation}\label{eq:R_np}
    G = \sigma\left\lbrack \frac{4l}{\pi d^{2}} + \frac{1}{d} \right\rbrack^{- 1}
\end{equation}

\subsection{Simulated electrical
measurements}\label{simulated-electrical-measurements}

To predict the sensing capabilities of nanopores integrated in
microfluidics, we simulated noise and translocation signals using an
electrical circuit model and dimensions of a model nanopore.

\subsubsection{Estimation of power spectral
density}\label{estimation-of-power-spectral-density}

The power spectral density (PSD, \(S_{i}\)) of a nanopore measurement
was estimated as the sum of five components \citep{tabard-cossa_instrumentation_2013, fragasso_comparing_2020}.

Flicker noise was calculated by

\[S_{1/f}(f) = \frac{\alpha_{H}I^{2}}{N_{\text{c}}f^{\beta}}\]

from the Hooge parameter \(\alpha_{H}\), current \(I\), number of charge
carriers \(N_{\text{c}}\), frequency \(f\) and empirical exponent
\(\beta\).

Thermal noise was

\[S_{\text{thermal}}(f) = \frac{4k_{\text{B}}T}{R}\]

with the Boltzmann constant \(k_{\text{B}}\), temperature \(T\) and
nanopore resistance \(R\) \citep{Kowalczyk2011a}.

Shot noise was calculated as

\[S_{\text{shot}}(f) = 2Iq\]

with the current \(I\) and effective charge of current carrying species
\(q\).

Dielectric noise was calculated as

\[S_{\text{dielectric}}(f) = 8\pi k_{\text{B}}TC_{\text{chip}}Df\]

with the chip capacitance \(C_{\text{chip}}\), dielectric loss \(D\).

Capacitive noise was calculated as

\[S_{\text{capacitance}}(f) = 4\pi^{2}C_{\text{total}}^{2}v_{n}^{2}f^{2}\]

with the total capacitance \(C_{\text{total}}\) and input voltage noise
\(v_{n}\).

\subsubsection{Calculation of time series from power spectral
density}\label{calculation-of-time-series-from-power-spectral-density}

The relationship between the PSD \(S_{i}\) and the Fourier transform
(\(\mathcal{F)}\) \citep{the_mathworks_inc_power_nodate} is

\[S_{i}(f) = \frac{\left| \mathcal{F}\left( I(t) \right) \right|^{2}}{Nf_{\text{s}}}\]

where \(N\) is the number of points in the time signal, and
\(f_{\text{s}}\) is the sample rate. The Fourier transform of the time
signal can then be found by

\[\mathcal{F}\left( I(t) \right) = \sqrt{N S_{i}(f)  f_{\text{s}}}\]

As the PSD lacks phase information, we generate a random phase $\theta$ 
for each frequency, with $\theta = e^{ix}$ and $x$ having a random value between 0 and $2\pi$.


%

To produce a current-time series, we then take the inverse real fast Fourier transform of the product of $\mathcal{F}$ and $\theta$. This signal has a noise profile matching the original PSD.



\subsubsection{Current blockade estimation using
SPICE}\label{current-blockade-estimation-using-spice}

Simulations of electrical circuit model for the nanopore were performed
using LTSPICE XVII and python 3.11 using the circuit from \ref{fig:circuit_blockade}.
The blockade was modeled as the addition of a resistance $\Delta R$ in equation \ref{eq:R_np} according to \citet{Kowalczyk2011a}:

\[d_{\rm blockade} = \sqrt{d_{\rm pore}^{2} - d_{\rm DNA}^{2}}\]

and the diameter of double-stranded DNA of 2.2~nm.

\section{Results}\label{results}

We present results showing that we could produce nanopore arrays and
measure integrated nanopores. We also present calculations predicting
the reduced noise of nanopores integrated by our methods.

\subsection{Fabrication and stability}\label{fabrication-and-stability}

\begin{figure}
    \centering
    \includegraphics{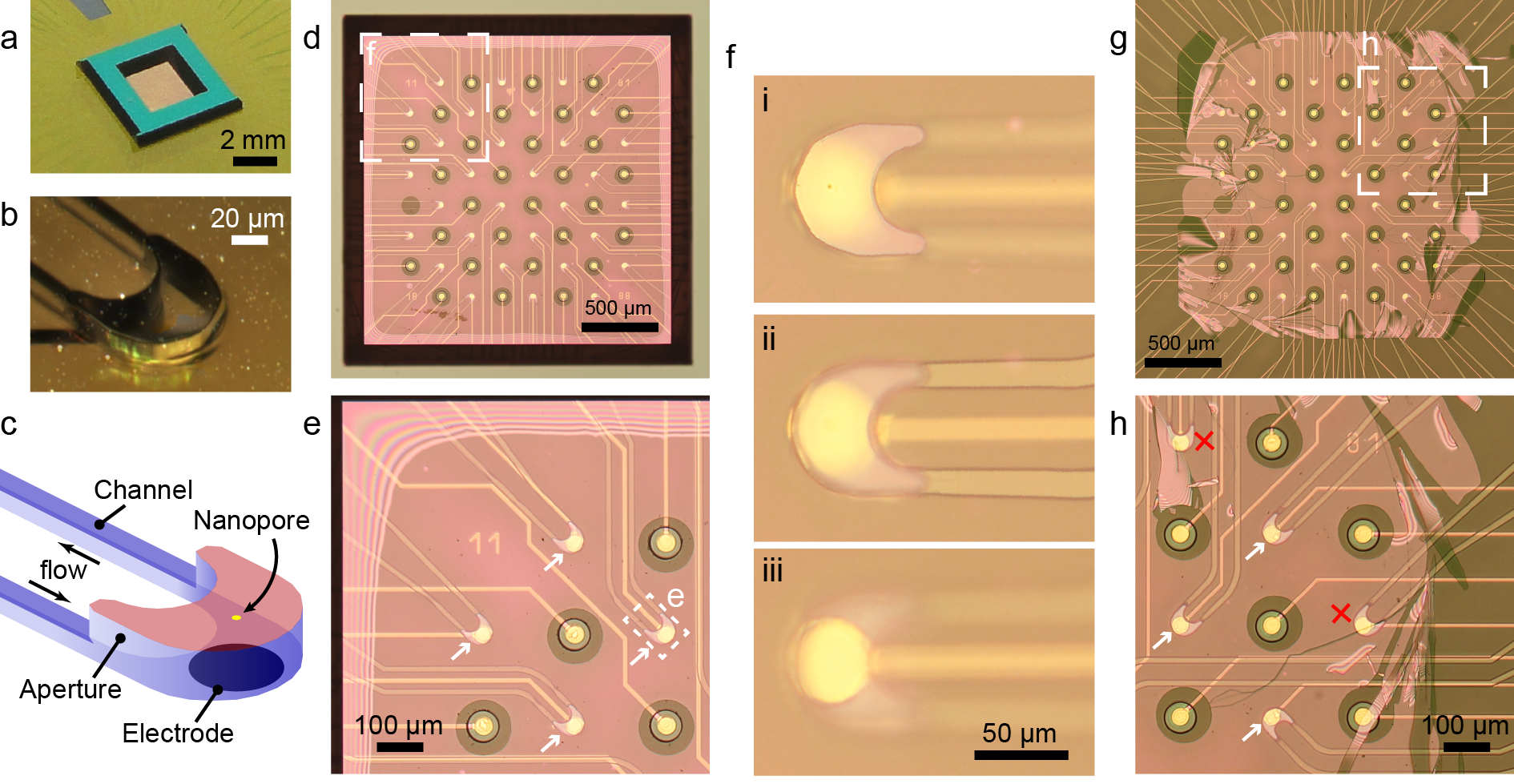}
    \caption{Nanopore array devices. \textbf{a:~}Photo of a nanopore array
with a 2×2~mm² membrane (in a 5×5~mm² chip) bonded to a microfluidic
network. \textbf{b:~}Oblique microscope photo (focus stack) of a single
nanopore site. \textbf{c:~}Oblique illustration of the aperture/nanopore
region of a microfluidic channel. Channel: blue. Membrane: red.
Nanopore: yellow. Electrode: black. \textbf{d,~e:}~Microscopic images of
the nanopore array region. The silicon nitride membrane is 2x2~mm². The
etched slopes of the silicon chip appear black. Interference patterns
indicate the unbonded region at the perimeter of the membrane, while the
dark region (ca. 1.9x1.9~mm²) is homogeneously bonded to the
microfluidic layer. In (e), white arrows indicate four free-standing SiN
membranes with nanopores. \textbf{f:~}Magnification of a single channel
with focus at (i) the nanopore membrane (the location of the nanopore
appears as a dark spot), (ii) the microfluidic channel and (iii) the
microelectrode. \textbf{g,~h:}~Microscopic images of a membrane which
fractured during bonding. Fracture at the perimeter of the membrane
detached the silicon frame. In comparison to the bonded regions,
unbonded regions appear darker or show bright interference patterns. In
(h), white arrows indicate three undamaged free-standing membranes
(white arrows) and red x's indicate two damaged regions.}
    \label{fig:devices}
\end{figure}

Our design (Figure~\ref{fig:devices}) had 30 out-and-back microfluidic channels
extending from peripheral access ports to a central nanopore region. In
the nanopore region, microfluidic channels had a width of 10~µm and a
minimum distance between channels of 40~µm. Peripherally, the channel
width increased to 50~µm to reduce hydraulic resistance. Channels had a
height of 10~µm and were enclosed by the 20~µm-thick aperture layer.
Neither collapse of the aperture layer nor bulging under pressure of up
to 6~bar was observed. The aperture layer defined a crescent-shaped
aperture (35~µm across) in each channel (Figure~\ref{fig:devices}f). The apertures'
crescent shape and low height/width ratio were intended to reduce the
risk of trapping bubbles (Figure~\ref{fig:devices}b,c). When filled with water, no air
bubbles were observed in the apertures.

When filled with water, we measured hydraulic resistances for single
channels on the order of 15~kPa·s/nL, which matched predicted values.
That is, pressure of 15~kPa would produce a flow rate of 1~nL/s.
Pressure of 100~kPa (1~bar) would produce a flow of 7~nL/s or
400~nL/min. The volume of each microfluidic channel was on the order of
10~nL.

Nanopores milled by a 10~pA gallium beam for 30~s had diameters of
\textless200~nm in 500~nm-thick membranes. The hydraulic resistance of
such a nanopore is on the order of 10\textsuperscript{5}~kPa·s/nL.
Alignment marks to enable optical alignment for bonding were also milled
by the gallium beam.

Dry bonding of silicon nitride membranes to the microfluidic apertures
formed a strong bond with a uniform appearance across most of the 2~mm membrane, except for unbonded peripheral regions (\ref{fig:devices}d,e).
Alignment precision was approximately 10~µm. During application of up to
6~bar for flow in the microfluidic channels, the membrane at the
aperture (at the halfway point of the channel) should have experienced a
pressure of 3~bar. No detachment was observed. Higher pressures were not
evaluated.\footnote{One exception: in an early experiment, we applied
  pressure by hand with a 1~mm syringe. The flow rate was high and
  uncontrolled, and generated extreme pressure which detached the
  membrane.}

In our bonding process, we manually brought the membrane into contact
with the top of the aperture layer, using a vertical displacement lever
on the FINEPLACER bonder and relying only on unmagnified visual
observation from the side. Unsurprisingly, this process produced varying
results. About half of bonded membranes were damaged, ranging from
fracture around the perimeter of the membrane (which allowed removal of
the Si frame) to shattering across the entire membrane. Unbroken
membranes showed homogeneous bonding across the 1.9~mm region contacted
directly by manipulator (Figure~\ref{fig:devices}d). Interference patterns at the
perimeter of the membrane and under the silicon frame (when viewed from
the bottom) showed that the frame itself was not bonded.

The broken membranes however further demonstrated the strength of the
bond. A membrane which shattered during bonding showed good adhesion
even with high pressure applied to an aperture with a crack within
20~µm (\ref{fig:devices}g,h).

We observed no deterioration of the microfluidic channels or membranes
when exposed to water for at least 2 weeks, neither after rinsing with
isopropanol, ethanol or acetone or blowing with pressurized air. An
ultrasonic bath shattered the free-standing regions of 500~nm-thick
membranes.

\subsection{Sixty-channel connector and fluidic
handling}\label{sixty-channel-connector-and-fluidic-handling}

\begin{figure}
    \centering
    \includegraphics{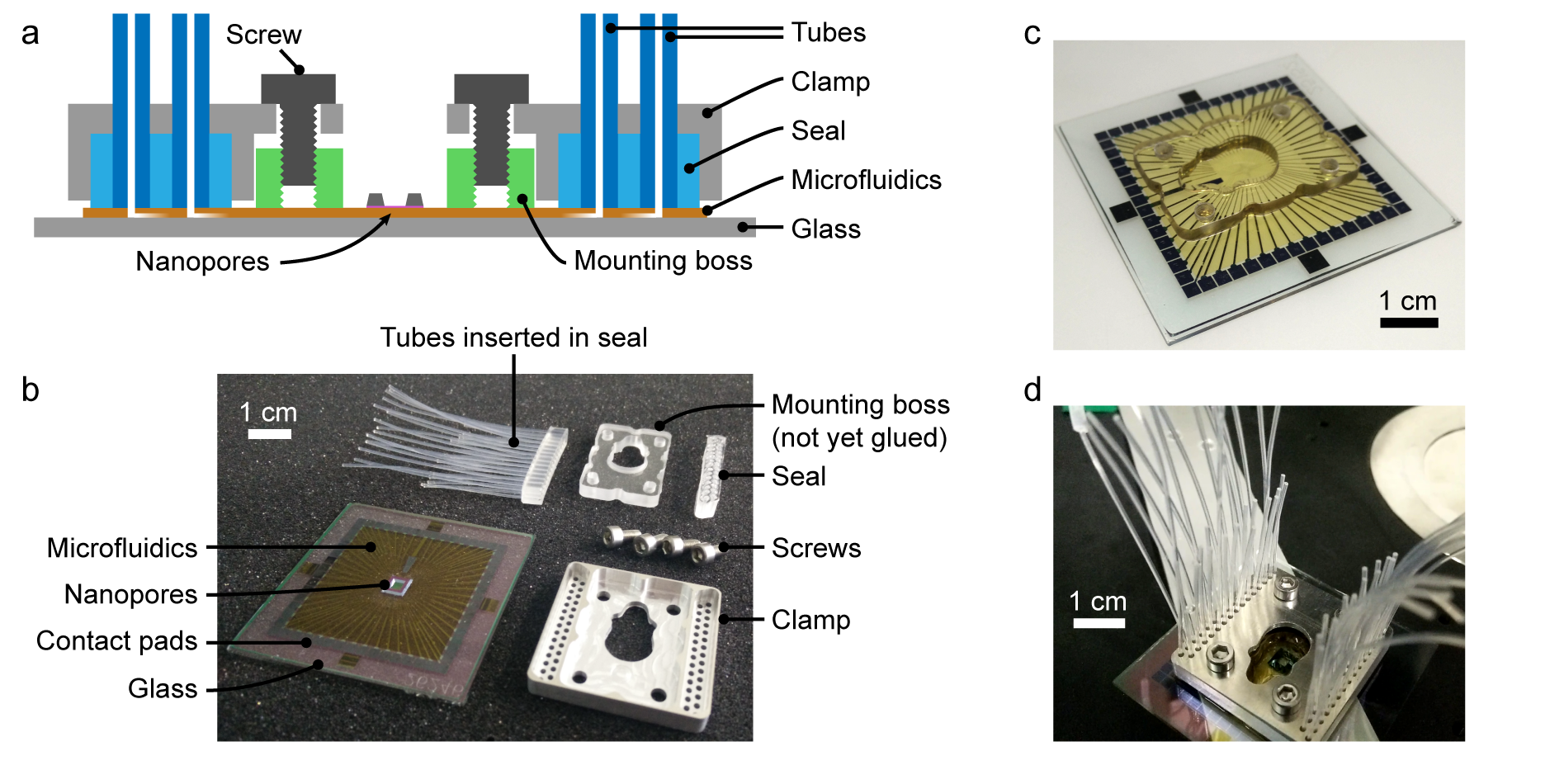}
    \caption{Microfluidic connector for the nanopore array. a:~Schematic
cross-section. b:~Nanopore array device and disassembled connector parts.
c:~Nanopore array with glued mounting boss. d:~Nanopore array assembled
with connector. In the top left, a thicker silicone tube is connected to
a single tube.}
    \label{fig:connector}
\end{figure}

The connector (Figure~\ref{fig:connector}) included two silicone seals, each capable of
clamping 30~tubes (with a pitch of 2~mm in two offset rows of 15), held
by an aluminum clamp. A tight fit was achieved by tubes with an outer
diameter of 0.8~mm (fluorinated ethylene propylene, 75~µm inner
diameter, IDEX H\&S, Wertheim-Mondfeld, Germany). A mounting boss,
permanently glued to each microfluidic substrate (\ref{fig:connector}c), ensured precise
alignment of the 60 tubes. The inner well of the boss defined the cis
compartment for all nanopores.

The 60-channel microfluidic connector formed a reliable connection to
the microfluidic chip. A good seal of the silicone seals against the
transparent substrate was visually observed from below. We observed an
absence of leakage by applying air pressure of up to 600~kPa while
submerging the connector and chip under water.

\subsection{Electrical measurements}\label{electrical-measurements}

\begin{figure}
    \centering
    \includegraphics{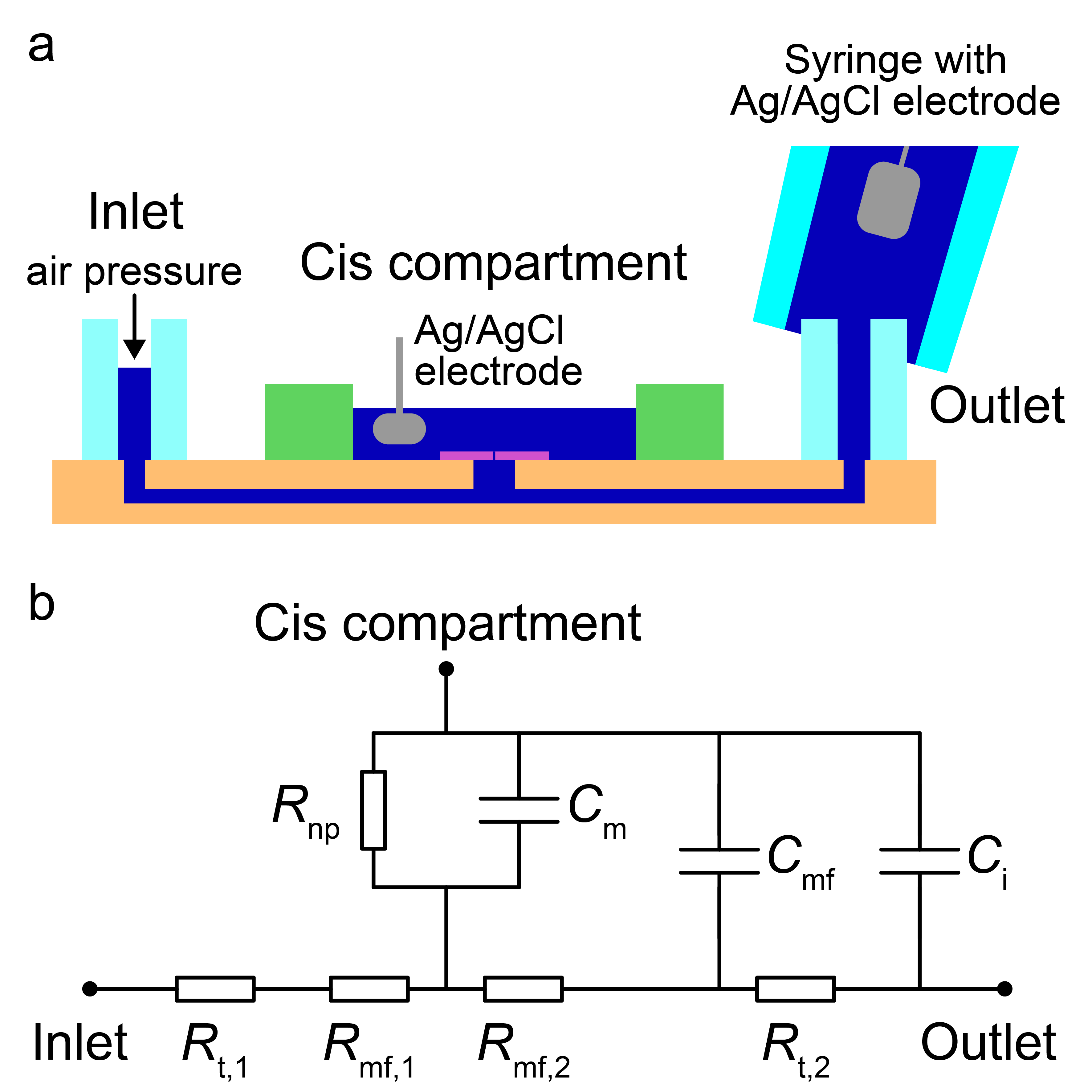}
    \caption{Set-up for measurements of nanopore arrays. a:~Simplified
schematic, not to scale. Measurements were made between Ag/AgCl
electrodes in the cis compartment and the outlet. The outlet tube was
loosely inserted into a syringe containing an electrode. Filling of the
microfluidic channel was controlled by air pressure at the inlet tube.
b:~A circuit model of a single microfluidic channel and nanopore
($R_{\rm np}$). The microfluidic channel is divided into two
components ($R_{\rm mf,1}$ and $R_{\rm mf,2}$) on either
side of the nanopore; in our design, these components are roughly equal
when both are filled. The resistance of the inlet and outlet tubes are
$R_{\rm t,1}$ and $R_{\rm t,2}$, respectively. The inlet
tube is connected to pressurized air ($R_{\rm t,1} \rightarrow \infty$)
and therefore measurements are made at the outlet. Importance
capacitances are those of the membrane ($C_{\rm m}$, pink in
panel a), the microfluidic channel (C\textsubscript{mf}, determined by
the length of the channel in contact with the cis compartment), and the
internal capacitance of the measurement electronics ($C_{\rm i}$).}
    \label{fig:circuit}
\end{figure}

\begin{figure}
    \centering
    \includegraphics{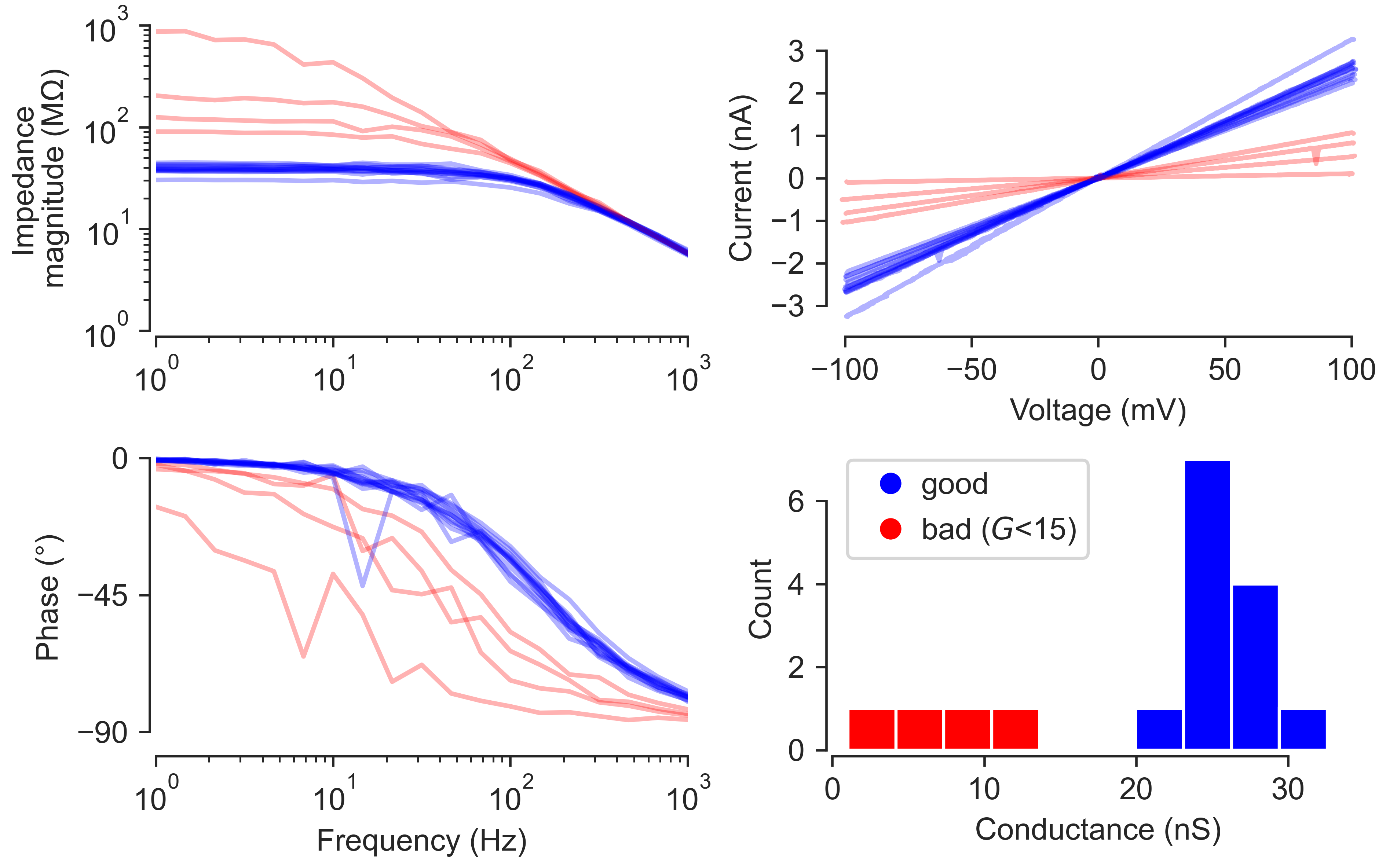}
    \caption{Electrochemical measurements of a nanopore array. Seventeen
nanopores were measured, of which 13 (blue) exhibited expected
conductances which included the series resistance of the microfluidic
channel. Four nanopores exhibited low conductances (red). Left: Bode
plots of impedance measurements. Right: Current--voltage plots and a
conductance histogram.}
    \label{fig:measurements}
\end{figure}

The setup and its circuit model are illustrated in Figure~\ref{fig:circuit}, with
estimated values in Table~\ref{tab:circuit}. Figure~\ref{fig:measurements} shows measurements from a
representative device. Eighteen nanopores were measured and showed
expected resistive behavior at low frequencies. Fourteen showed similar
conductances of 25~±~3~nS (resistance 40~M$\Omega$), while four channels had
lower conductances which we attribute to partial blockages in the
microfluidic channels or tubing. Twelve nanopores were not measured due
to blockages which prevented filling of the microfluidic channels.

\begin{table}[]
    \caption{Estimated values for circuit elements of the integrated
nanopores. Sources: relative permittivity of SiN \citep{Piccirillo1990}; relative permittivity of ADEX is assumed to be similar to SU-8 \citep{ghannam_dielectric_2009}; conductivity of PBS measured to be 1.5~S/m.}
    \centering
    \begin{tabular}{ccp{7cm}}
\toprule
\text{Element} & \text{Value} & \text{Details} \\ \midrule
$R_{\rm np}$         & 14 M$\Omega$             & Diameter 200 nm, length 500 nm \\
$C_{\rm m}$            & 0.3 pF                 & 500 nm-thick SiN membrane, relative permittivity of 7, aperture area of 2500 µm² \\
$C_{\rm mf}$         & 0.2 pF                 & 20 µm-thick ADEX, relative permittivity of 2.85, channel area of 0.2 mm² \\
$C_{\rm i}$            & 20 pF             & Parasitic capacitance \\
$R_{\rm t,1}$, $R_{\rm t,2}$  & 6 M$\Omega$            & Diameter 75 µm, length 4 cm \\
$R_{\rm mf,1}$, $R_{\rm mf,2}$ & 15 to 24 M$\Omega$    & Channel lengths 15–-21 mm, height 15 µm, widths 10-–50 µm \\
$R_{\rm np} + R_{\rm mf,2} + R_{\rm t,2}$ & 35 to 44 M$\Omega$ & \\ \bottomrule
    \end{tabular}
	\label{tab:circuit}
\end{table}


\subsection{Simulated electrical
measurements}\label{simulated-electrical-measurements-1}

\begin{figure}
    \centering
    \includegraphics{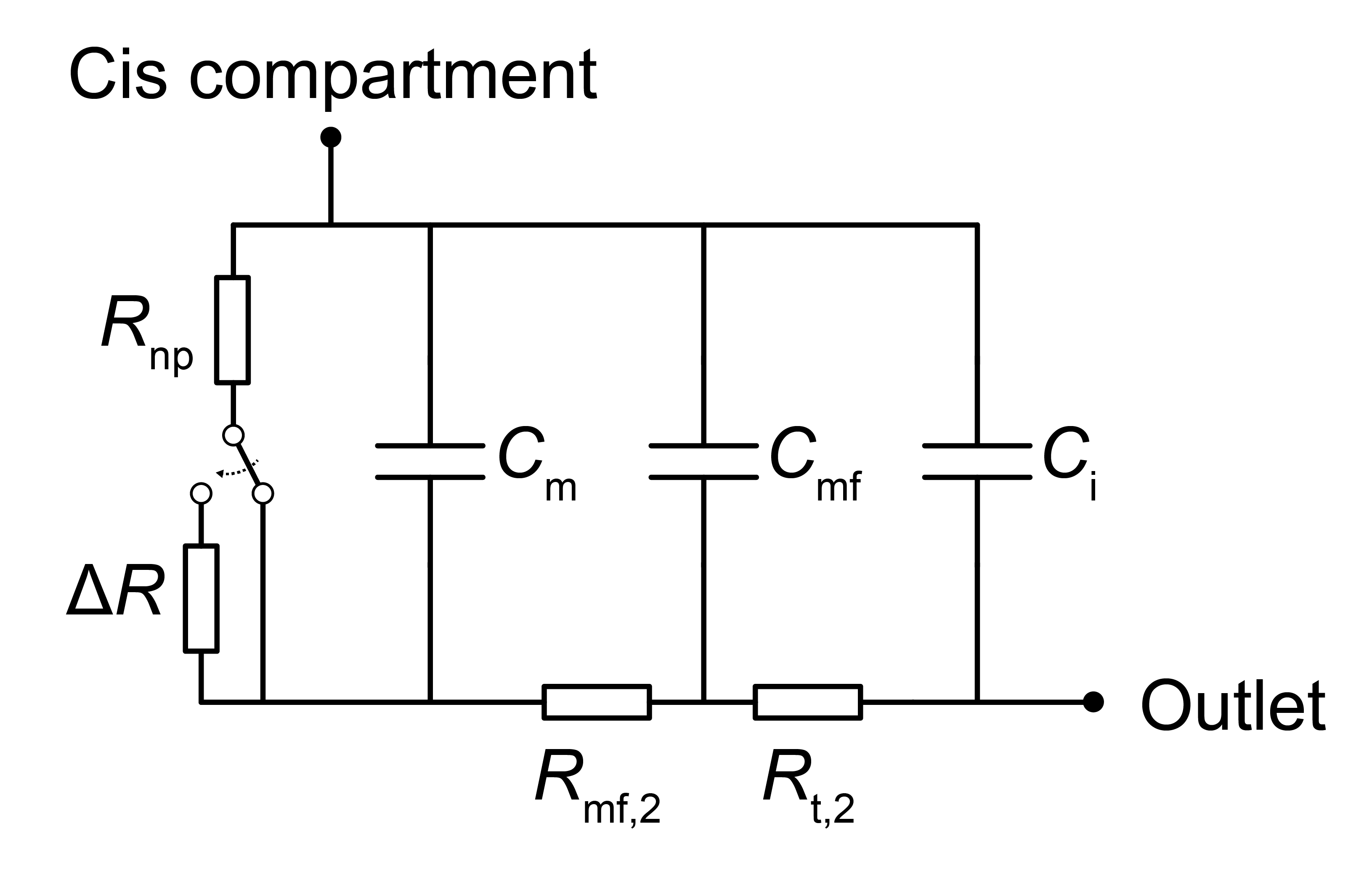}
    \caption{Circuit for simulations.}
    \label{fig:circuit_blockade}
\end{figure}

Our experimental demonstration used large nanopores, whereas most
sensing applications require diameters far smaller than the 20~nm that is possible using a gallium FIB. To evaluate the
effect of microfluidic integration on nanopore sensing, we performed
simulations using a model circuit (Figure~\ref{fig:circuit_blockade}) and a model nanopore with a
diameter of 10~nm. The resistor $\Delta R$ represents the change of
resistance during a translocation. Full details are in Table~\ref{tab:simulations}.

\begin{table}[]
    \caption{Values for simulations.}
    \centering
    \begin{tabular}{ccp{7cm}}
\toprule
\text{Element} & \text{Value} & \text{Details} \\ \midrule
$\sigma$ & 10.5 S/m & 1~M KCl \\
$R_{\rm np}$ & 130.8~M$\Omega$ (open) & Open diameter 10~nm,
length 100~nm (Equation 3) \\
$R_{\rm np}+\Delta R$ & 137.2 M$\Omega$ (blocked) &
Double-stranded DNA diameter of 2.2~nm \\
$C_{\rm m}$ & 1.55~pF & 100~nm-thick SiN membrane,
relative permittivity of 7, aperture area of 2500~µm² \\
$C_{\rm mf}$ & 0.2~pF & 20~µm-thick ADEX, relative
permittivity of 2.85, channel area of 0.2~mm² \\
$C_{\rm i}$ & 1 pF & Internal capacitance of the
Axopatch 200B \citet{molecular_devices_llc_axon_2012} \\
$R_{\rm t,2}$ & 172 k$\Omega$ & Diameter 75~µm, length 8~mm, $\sigma$ \\
$R_{\rm mf,2}$ & 1.905 to 13.33 M$\Omega$ & Channel lengths
15--21~mm, height 15~µm, widths 10--50~µm. 1.905 M$\Omega$ used in most
simulations unless otherwise noted. \\
$R_{\rm np}+R_{\rm mf,2}+R_{\rm t,2}$ & 138 to 150~M$\Omega$ & \\
$f$ & 0.1 to 4 MHz & \\
$f_{\rm c}$ & 100 kHz & Second-order Bessel filter \\
$T$ & 298 K & \\
$V$ & 500 mV & \\
$V_{\rm n}$ & 3 nV/Hz\textsuperscript{-1} & \citet{molecular_devices_llc_axon_2012} \\
$D$ & 10\textsuperscript{‑3} & \citet{pan_measurement_2023} \\
\bottomrule
    \end{tabular}
	\label{tab:simulations}
\end{table}

\begin{figure}
    \centering
    \includegraphics{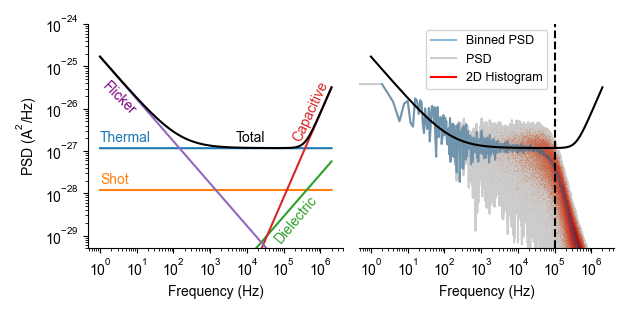}
    \caption{Theoretical PSD of a microfluidic-integrated nanopore. Left:
Total noise was calculated from contributions of flicker, thermal, shot
and capacitive noise. Dielectric noise was negligible. Right: A more
realistic PSD was calculated from a theoretical low-pass filtered
current trace. The dashed vertical line indicates the 100~kHz filter.
The PSD (grey) is additionally plotted using logarithmic binning (blue)
and a 2D histogram (red). The total PSD from the left panel is overlaid
(black).}
    \label{fig:PSD}
\end{figure}

First, we calculated the noise expected in a nanopore measurement.
Figure~\ref{fig:PSD} shows the noise PSD, including the constituent components and
total (left). Notably, this theoretical PSD is not limited to any
bandwidth. To produce a more realistic PSD, the total PSD was
transformed into a time-domain current trace, filtered using a 100~kHz
low-pass filter, and a new PSD was calculated (Figure~\ref{fig:PSD}, right).

\begin{figure}
    \centering
    \includegraphics{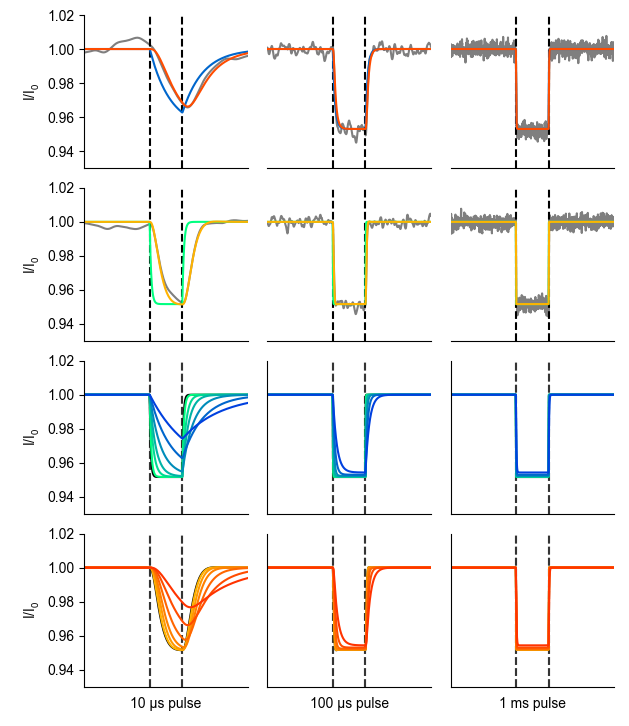}
    \caption{Effect of series resistance on translocation measurements.
Microfluidic integration can filter recordings due to addition of a
series resistance. First row: Simulated pulses filtered due to
integration in a high resistance (8~M$\Omega$) microfluidic channel (blue),
additionally filtered by the 100~kHz lowpass filter of the AxoPatch 200B
(red) and with simulated noise (grey). Second row: Simulated pulses
filtered due to integration in a low resistance (0.1~M$\Omega$) microfluidic
channel (green), additionally filtered by the 100~kHz lowpass filter of
the AxoPatch 200B (orange) and with simulated noise (grey). Third row:
Simulated pulses filtered due to integration in microfluidic channels
having resistances from 0.1~M$\Omega$ (green) to 8~M$\Omega$ (blue). Fourth row: The
traces from the third row, additionally filtered by the 100~kHz lowpass
filter of the AxoPatch 200B. Simulated pulses have durations of 10~µs,
100~µs and 1~ms (left to right).}
    \label{fig:blockades}
\end{figure}

Next, we examined the shape of ionic current blockades. Blockade depth
was calculated for double-stranded DNA. Blockade durations of 10~µs,
100~µs and 1~ms were considered. Simulations predict that microfluidic
integration will filter the recordings due to additional series
resistance (Figure~\ref{fig:blockades}). Specifically, the simulations varied
$R_\text{mf,2}$ from 0.1 to 8~M$\Omega$. The high resistance
corresponds to the use of an external electrode which accesses the nanopore
via a microfluidic channel. The low resistance corresponds to a device in
which an electrode is integrated in the proximity of a nanopore.

The effect of series resistance is contrasted in the first two rows,
showing blockades with series resistances of 8~M$\Omega$ and 0.1~M$\Omega$,
respectively. Blockades with a 1~ms duration are well resolved. However,
the 100~µs blockade in the top row is distorted due to the high series
resistance. The effect is stronger for 10~µs blockades: the unfiltered
blockade with low resistance (green) is nearly rectangular, while the
unfiltered blockade with high resistance (blue) is significantly
distorted. The effect of the 100~kHz low-pass filter becomes significant
for such short pulses (orange and red).

In both rows, grey traces include noise simulated using SPICE. Noise was
calculated for an open pore and does not consider the slight changes in
noise during a blockade.

The third and fourth rows of Figure~\ref{fig:blockades} illustrate this effect for a range
of resistances, first unfiltered and then with a 100~kHz low-pass
filter.

These results demonstrate that close integration between electrodes and
nanopores will be important for high bandwidth measurements. Moreover,
these results show that the filtering caused by a high series resistance
is qualitatively similar to a low-pass filter. Interpretation of
measurements using integrated nanopore devices must consider both
effects.

\section{Discussion}\label{discussion}

Here, we contrast our fabrication methods with previous work. We then
focus on limitations of our devices (especially handling and usability).
We also discuss noise.

\subsection{Fabrication of nanopore array
devices}\label{fabrication-of-nanopore-array-devices}

Microfluidic integration of single nanopores or arrays of up to eight
nanopores has previously been achieved using PDMS microfluidics \citep{Jain2013, jain_microfluidic_2017, Tahvildari2015, Tahvildari2016}. Thin film microfluidics
as reported here would be challenging in PDMS, due to its flexibility.
Furthermore, its ability to absorb small molecules makes it undesirable
\citep{toepke_pdms_2006}. Therefore, we were motivated to integrate
nanopores in epoxy photoresists to achieve thin layers and high
resolution, thereby enabling higher nanopore density. Epoxy photoresists
have proven suitable for nanopore sensing applications and use with
Ag/AgCl microelectrodes in microelectrode cavity array (MECA) devices
\citep{Baaken2008}. ADEX is chemically similar to SU-8, and both can
withstand solvents including ethanol, isopropanol, and acetone.
Photolithographic patterning of epoxy resists could be extended to
wafer-scale processes and CMOS wafers.

A challenge was to develop clean enclosed channels covered by ADEX.
Without rinsing, residue remained in the channels. And without
blow-drying, solvent remained in the channels. In our design, the
central apertures enabled manual rinsing with fresh solvent followed by
blow-drying with nitrogen, which achieved clean, dry channels. However,
such manual rinsing and drying will not be possible when scaling to
wafer-scale production with many chip replicates per wafer.

Previous reports of integration in PDMS microfluidics have contacted
nanopore membranes from both sides, so that chemical bonding is
additionally physically supported. A single-sided bonding method makes
interfacing simpler but cannot rely on physical support. Therefore, we
developed a dry bonding method based on amine--epoxide polyaddition.
Similar methods have previously been reported for bonding of SU-8 and
PDMS \citep{Zhang2009, Zhang2011, Watts2012}, by exploiting
residual epoxy groups in SU-8 and functionalization of PDMS with primary
amine groups. The soft elastomeric nature of PDMS enabled good contact
with comparably rigid SU-8. We believe that our successful bonding
results were possible by supporting the flexible yet fragile SiN
membrane on a soft silicone manipulator, and softening of the ADEX layer
at elevated temperatures (before the later hard-bake to complete
cross-linking). We expect that our roughly 50\% chance of fracturing the
membrane can be decreased with better process control, although this
must be further investigated for thinner membranes.

In our bonding process, force could be roughly controlled with a
weighted arm. Several membranes bonded with a force of 5~N shattered. At
forces of 2--3~N, about half of the membranes shattered while the other
half remained undamaged. We expect that the variability could be
improved by magnified visualization during the bonding process or by
improved force sensing and control. In some cases, the membrane was
bonded while the Si frame cleanly broke off -- similar results can be
intentionally achieved by perforation of the membrane \citep{Jain2013}. In cases where cracks in the SiN membrane were near
(\textasciitilde20~µm) but not contacting an aperture, we applied
pressure of several bar and observed no weakness. We also expect that bonding may be achieved much faster than the 45~min used here.

For dry bonding of nanopore membranes, we chose the monofunctional
APDMES, because common trifunctional silanes (such as
aminopropyltriethoxysilane, APTES) can form thick layers \citep{Vandenberg1991} 
which could block nanopores. The SiN membranes shatter in an
ultrasonic bath which is often used to remove loosely bonded molecules.
Electrical measurements confirmed that nanopores were unblocked after
APDMES coating and bonding, although an organosilane coating does reduce
nanopore size \citep{Wanunu2007}.

Further developments of adhesive bonding may be necessary for other
membrane materials. New methods capable of orthogonally
functionalization of the bonding regions of a membrane and the interior
nanopore walls may also be needed for specific sensing capabilities.

We have also bonded nanopore arrays using a liquid adhesive
\citep{Kentsch2006}. The liquid adhesive risked filling the channels
or nanopores, and even in successful cases extended a meniscus 10-20~µm
along the membrane into the apertures. In comparison, dry bonding caused
no deformation of the microfluidic structures.

We produced large nanopores using a gallium FIB, which readily
patterns nanopores over large areas but cannot achieve nanopores smaller
than 20~nm. Our nanopore arrays were patterned in a 1.4~mm square, which
is beyond the field of view of conventional transmission electron
microscopes. Milling with a helium ion beam would enable large arrays
with nanopore sizes competitive with electron beam milling. We also
produced alignment marks using the gallium beam, which would be
impractical with an electron or helium beam, and should rather be
produced lithographically. Rather than milling nanopores, controlled
breakdown avoids the need for precise bonding alignment \citep{Tahvildari2015}. Nanopore fabrication by reactive-ion etching may be
beneficial for scale-up \citep{bai_fabrication_2014, martens_nanopore-fet_2022}.

\subsection{Electrical properties of nanopore
arrays}\label{electrical-properties-of-nanopore-arrays}

In our work, we estimated (Table~\ref{tab:circuit}) that the capacitance of a suspended
membrane was 0.3~pF, and that the microfluidic channel added a
capacitance of 0.2~pF. These capacitances must be minimized to limit
current noise. The 20~pF internal capacitance of the potentiostat
prevented better analysis of these capacitances.

The 500~nm thickness of our membranes limits comparison with other
nanopores. We estimate a 20~nm‑thick membrane suspended over the
microfluidic apertures would have a capacitance of about 10~pF -- much
higher than desired. Good solutions would be to have thinned regions on
a thick membrane (which would also facilitate handling and bonding) and
to reduce the aperture size. We have produced apertures as small as 6~µm
in 20~µm-thick ADEX, although such high aspect ratio features may trap
bubbles and are sensitive to process fluctuations. For smaller aperatures, we expect good
results would be possible using 5~µm-thin ADEX.

The effect of the high resistances of the microfluidic channels and
tubes on sensing capabilities should be investigated in future work.
Integration of Ag/AgCl microelectrodes near the nanopores and the use of
3- or 4-electrode systems could be considered.

\subsection{Fluidic handling}\label{fluidic-handling}

Future nanopore devices should be simple and easy to use, however this
was not achieved in this work. Although we solved the challenge of
connecting 60 tubes to a microfluidic chip, handling 30 channels remains
challenging and flow control was done serially in all channels. Neither
syringe pumps nor pressure-driven pumps are readily available in
30-channel formats.

Each channel required two tubes, as interfacing to a nanopore generally
requires a channel having both an inlet and an outlet. Nanopores
themselves have enormous hydraulic resistances (Equation 1), so that
filling or rinsing from a single inlet through the pore itself is
impractical. Rinsing our microfluidic channels was slow (minutes per
microliter), however rinsing through a nanopore would be much slower
(weeks or months per microliter).

Our serial process for filling channels and measuring impedance spectra
required about 12~min per nanopore, or \textgreater6~hours for the
array. The low volume of the microfluidic channel (\textasciitilde10~nl)
results in a residence time of water of about 1~s (at 1~bar). This
suggests that handling 30 channels in a reasonable timeframe could be
possible if a technical solution were developed. Directly pipetting into
each channel would be convenient, however air cushioned pipettes apply
low pressure and positive displacement pipettes would risk damaging the
membranes due to high pressures. We're not aware of pressure-controlled
pipettes which would make 30 channels possible (although laborious).
Scaling would quickly become absurd. Overall, we believe that any
devices with similar or higher numbers of nanopores require new
microfluidic concepts to integrate multiple nanopores per microfluidic
channel. An example would be multiplexing using microfluidic control
valves \citep{Tahvildari2016}, although using materials other than
PDMS would be desirable \citep{roy_thermoplastic_2011}. Future work could
investigate on-chip sample processing \citep{varongchayakul_solidstate_2018}
before read-out using nanopore arrays.

\subsection{Noise of integrated
nanopores}\label{noise-of-integrated-nanopores}

Other groups have shown a reduction in noise using more highly
integrated systems for measuring nanopores \citep{rosenstein_integrated_2012}.
Our results support this, and furthermore highlight the detrimental
effect that electrical resistance of microfluidic channels may have.

The current noise PSD (Figure~\ref{fig:PSD}) showed that the low capacitance
realized by microfluidic integration will minimize capacitive or dielectric
noise at a bandwidth up to 100~kHz. Previous efforts have reduced
capacitance by manually painting insulating silicone on nanopore
membranes \citep{Balan2014}. Compared to manually painting the
insulating layer, the use of photolithographic microfabrication and dry
bonding allows for a scalable and reproducible method for reducing the
current noise of nanopores.

The presented microfluidic chip demonstrates the potential for low-noise, high-bandwidth measurements (well beyond 100 kHz). While SPICE simulations indicate that high series resistance in the microfluidic channels could impact signal quality, this challenge can be addressed with optimized design strategies. Reducing the electrical resistance of the microfluidic channels will enhance the performance of the nanopore device by minimizing signal attenuation at higher frequencies. Achieving a resistance below 1 MΩ, for instance, would enable resolution of short 10 µs blockades. Moving forward, integrating microelectrodes close to the nanopores, along with the use of CMOS chips for on-site amplification and digitization, could further advance the capabilities of integrated microfluidic nanopore arrays

Beyond achieving low noise in high bandwidth measurements, our results highlight challenges for resolving short signals.
SPICE simulations of ionic current blockades (Figure~\ref{fig:blockades}) demonstrate that the high series resistance introduced by a microfluidic channel between a nanopore and an electrode increases the RC time constant of the nanopore
device, leading to signal attenuation at higher frequencies. Integrated
nanopore arrays must minimize this resistance to enable high bandwidth measurements. For example, a
resistance less than 1~M$\Omega$ would be necessary to resolve 10~µs blockades.
The ideal solution would be to integrate microelectrodes in direct
proximity to the nanopores. Beyond the use of integrated microfluidic
nanopore arrays on glass substrates, further improvement will be gained
by direct integration on CMOS chips for amplification and digitization.

\subsection{Limitations of this work}\label{limitations-of-this-work}

A limitation of our work is the use of thick membranes and large
nanopores. Our design required a grid of nanopores over an area of
1.4x1.4~mm². We could easily pattern these nanopores and optically
visible alignment marks using a gallium FIB. An array over such a large
area using a TEM would be challenging. Future work should integrate
alignment marks using RIE and produce nanopores using helium ion milling
which can achieve single-digit nanometer diameters.

Stability of nanopores remains a concern \citep{chou_lifetime_2020}. Different
membrane materials or coatings such as HfO\textsubscript{2} should
improve nanopore stability, but their integration using our methods
may require changes to bonding chemistry.

Our minimum microfluidic feature sizes of 10~µm were straightforward to
fabricate for this demonstration. Increasing resolution could easily
increase channel/nanopore density, however handling is currently the
limitation.

The integration of electrodes is another issue to be solved, and their
size may limit nanopore array density. Our design included electrodes
with a diameter of 30~µm (which we did not use for measurements),
similar to MECA chips for biological nanopores \citep{Baaken2008}.
Integrated electrodes for solid-state nanopores may require diameters 10
times larger. If using glass substrates, aligning the nanopore over a
window in the electrode would allow correlative optical measurements
\citep{ensslen_chip-based_2022}.

\section{Conclusions}\label{conclusions}

We have presented a microfluidic chip featuring the largest number of
individually-addressed solid-state nanopores to date. Our dry bonding
method established a stable chemical bond between silicon nitride
membranes and microfluidic networks, while enabling maximum process
flexibility by separating nanofabrication and microfabrication
processes. The method should be compatible with CMOS substrates
containing integrated microelectrodes and alternative nanopore
fabrication methods such as helium ion beam milling, reactive-ion etching or controlled
breakdown. Measurements confirmed the electrical isolation of the
integrated nanopores, and calculations demonstrated that integrated
nanopores can enable recordings with reduced current noise. A major
limitation is the impractical chip-to-world connection, which required
60 tubes for the 30 microfluidic channels. Simulations show that integration of electrodes near the nanopores will be necessary for high bandwidth measurements. Our fabrication method holds
promise when applied to new designs that improve fluidic interfacing and electrode integration.

\section{Acknowledgements}\label{acknowledgements}

This work was supported by the European Commission (7th Framework
Programme, Marie Curie Initial Training Network ``NAMASEN,'' contract n.
264872), the VW Foundation under their Experiment! Program, the State
Ministry of Economic Affairs, Labour and Tourism as part of the Forum
Gesundheitsstandort Baden-Württemberg (TechPat nano), and the Federal
Ministry of Education and Research (BMBF) in the project nanodiag\_BW
(03ZU1208BG).

\bibliographystyle{abbrvnat}
\bibliography{references}  






\end{document}